\documentclass{llncs}

\usepackage{amsmath,amsfonts,latexsym}

\newcommand{\cross}{\otimes}

\newcommand{\ket}[1]{\mbox{$|#1\rangle$}}

\def\ceil#1{\lceil#1\rceil}

\newcommand{\squash}[1]{\raisebox{0.04ex}[0pt][0pt]{\small$\textstyle #1$}}
\newcommand{\oosrt}{\squash{\frac{1}{\sqrt{2}}}}

\newcommand{\optprob}{\mbox{\smash{$\frac{1}{2} + 2^{-\ceil{n/2}}$}}}

\newcommand{\pbfrac}[2]{\mbox{$\mbox{}^{#1}\!/_{#2}$}}
\newcommand{\mypmod}[1]{~(\textup{mod}~#1)}

\begin{document}

\title{Multi-Party Pseudo-Telepathy}


\author{
Gilles Brassard\,%
\thanks{Supported in part by Canada's {\sc Nserc}, Qu\'ebec's {\sc Fcar},
the Canada Research Chair Programme,
and the Canadian Institute for Advanced Research.}
\and
Anne Broadbent\,%
\thanks{Supported in part by a scholarship from Canada's {\sc Nserc}.}
\and
Alain Tapp\,%
\thanks{Supported in part by Canada's {\sc Nserc} and Qu\'ebec's {\sc Fcar}.}
}

\institute{D\'{e}partement IRO, Universit\'{e} de Montr\'{e}al, 
C.P. 6128, succursale centre-ville, Montr\'{e}al (Qu\'{e}bec), Canada H3C 3J7
\\ \email{\{brassard,\,broadbea,\,tappa\}@iro.umontreal.ca}}

\maketitle

\sloppy


\begin{abstract}
Quantum entanglement, perhaps the most non-classical manifestation of
quantum information theory, cannot be used to transmit
infor\-mation between remote parties.
Yet, it can be used to \mbox{reduce} the amount of communication
required to process a variety of \mbox{distributed} computational tasks.
We speak of \emph{pseudo-telepathy} when quantum \mbox{entanglement}
serves to \emph{eliminate} the classical need to communicate.
In~earlier examples of pseudo-telepathy, classical protocols
could \mbox{succeed} with high probability unless the inputs
were very large.
Here we present a simple multi-party
distributed problem for which the \mbox{inputs} and \mbox{outputs} consist of
a single bit per player, and we~present a \mbox{perfect} quantum \mbox{protocol}
for~it. We prove that no classical protocol can \mbox{succeed} with a
probability that differs from \pbfrac{1}{2} by more than a fraction
that is exponentially small in the number of \mbox{players}.
This could be used to circumvent the detection loophole
in experi\-mental tests of nonlocality.

\end{abstract}

\section{Introduction}
\label{introduction}

It is well-known that quantum mechanics can be harnessed to reduce
the amount of communication required to perform a variety of distributed
tasks~\cite{survey}, through the use of either quantum communication~\cite{yao83}
or quantum entanglement~\cite{CB}.
Consider for example the case of Alice and Bob, who are very busy and
would like to find a time when they are simultaneously free for lunch.
They each have an engagement calendar, which we may think of as
\mbox{$n$--bit} strings $a$ and~$b$, where
\mbox{$a_i=1$} (resp.~\mbox{$b_i=1$}) means that Alice (resp.~Bob) is free for
lunch on day~$i$.  Mathematically, they want to find an index~$i$ such that
\mbox{$a_i=b_i=1$} or establish that such an index does not exist.
The obvious solution is for Alice, say, to communicate her entire calendar
to Bob, so that he can decide on the date: this requires roughly
$n$ bits of communication.  It turns out that this is optimal in the worst
case, up to a constant factor, according to classical information
\mbox{theory}~\cite{KS}, even when the answer is only required to be correct
with probability at least~\pbfrac{2}{3}.  Yet, this problem can be solved with
arbitrarily high success probability with the \mbox{exchange} of a number of
\emph{quantum} bits---known as \emph{qubits}---in
the order of~$\sqrt{n}$~\cite{aaronson}. 
Alternatively, a number of \emph{classical} bits in the order of
$\sqrt{n}$ suffices for this task if Alice and Bob share prior entanglement,
because they can make use of quantum teleportation~\cite{teleport}.
Other (less natural) problems demonstrate an \emph{exponential}
advantage of quantum communication, both in the error-free~\cite{BCW} and
bounded-error~\cite{raz} models.

Given that prior entanglement allows for a dramatic \emph{reduction} in the
need for classical communication in order to perform some distributed
computational tasks, it is natural to wonder if it can
be used to \emph{eliminate} the need for communication altogether.
In~other words, are there distributed tasks
that would be impossible to achieve in a classical world if the participants were
not allowed to communicate, yet those tasks could be performed without
\emph{any} form of communication provided they share prior
entanglement? The answer is negative if the result of the computation must
become known to at least one party, but it is positive if we are satisfied with the
establishment of nonlocal correlations between the parties' inputs and
outputs~\cite{BCT99}.

Mathematically, consider $n$ parties $A_1$, $A_2$,\ldots, $A_n$
and two \mbox{$n$-ary} functions $f$ and~$g$\@.
In~an \emph{initialization phase}, the parties are allowed to discuss
strategy and share random variables (in the classical setting)
and entanglement (in the quantum setting).  
Then the parties move apart and they are no longer allowed any form of
communication. After the parties are physically separated,
each $A_i$ is given some input $x_i$ and is requested to produce output~$y_i$.
We~say that the parties \emph{win} this instance of the game if
\mbox{$g(y_1,\,y_2,\ldots\,y_n)=f(x_1,\,x_2,\ldots\,x_n)$}.
Given an $n$-ary predicate $P$, known as the \emph{promise}, a protocol is
\emph{perfect} if it wins the game with certainty on all inputs 
that satisfy the promise, i.e.~whenever \mbox{$P(x_1,\,x_2,\ldots\,x_n)$}
holds. A~protocol is \emph{successful with probability} $p$
if it wins \emph{any} instance that satisfies the promise with probability
at least~$p$;
it~is successful in \emph{proportion} $p$ if it wins the game
with probability at least $p$ when the instance is chosen at random
according to the uniform distribution on the set of instances that satisfy
the promise.
Any protocol that succeeds with probability $p$ automatically
succeeds in proportion~$p$, but not necessarily vice versa.
In~particular, it is possible for a protocol that succeeds in
proportion $p>0$ to fail systematically on some inputs, whereas this
would not be allowed for protocols that succeed with probability~$p>0$.
Therefore, the notion of succeeding ``in~proportion'' is meaningful for
\emph{deterministic} protocols but not the notion of succeeding ``with
probability''.

We say of a quantum protocol that it exhibits
\emph{pseudo-telepathy} if it is perfect provided the parties
share prior entanglement, whereas no perfect classical protocol can exist.
The~study of pseudo-telepathy was initiated in~\cite{BCT99},
but all examples known so far allowed for classical protocols that
succeed with rather high probability, unless the inputs are very long.
This made the prospect of experimental demonstration of pseudo-telepathy
unappealing for two reasons.
\begin{itemize}
\item[$\diamond$] It would not be surprising for several runs of an imperfect
classical protocol to succeed, so that mounting evidence of a convincingly quantum
behaviour would require a large number of consecutive successful runs.
\item[$\diamond$] Even a slight imperfection in the quantum implementation would
be likely to result in an error probability higher than what can easily be achieved
with simple classical protocols!
\end{itemize}
In Section~\ref{quantum}, we introduce a simple multi-party
distributed computational problem for which the inputs and outputs consist of
a single bit per player, and we~present a perfect quantum protocol for~it.
We prove in Sections~\ref{classical}
and~\ref{probabilistic} that no classical protocol can succeed with a
probability that differs from \pbfrac{1}{2} by more than a fraction that is
exponentially small in the number of players.  More \mbox{precisely}, no
classical protocol can succeed with a probability better than
$\optprob$, where $n$ is the number of players.
Furthermore, we show in Section~\ref{loophole} that the success probability of
our quantum protocol would remain better than anything classically achievable,
when $n$ is sufficiently large, even if each player had imperfect apparatus that
would produce the wrong answer with probability nearly 15\% or no answer at all
with probability~29\%.
This could be used to circumvent the infamous \emph{detection loophole} in
experimental proofs of the nonlocality of the world in which we
live~\cite{massar}.

\section{A Simple Game and its Perfect Quantum Protocol}
\label{quantum}

For any $n \ge 3$, game $G_n$ consists of $n$ players.
Each player $A_i$ receives a single input bit $x_i$ and is
requested to produce a single output bit~$y_i$.
The players are promised that there is an even number of 1s
among their inputs. Without being allowed to communicate
after receiving their inputs, the players are challenged to
produce a collective output
that contains an even number of 1s if and only if the
number of 1s in the input is divisible by~4.
More formally, we require that
\begin{equation}\label{goal}
\sum_i^n y_i ~\equiv~
{\textstyle \frac12} {\sum_i^n x_i} \pmod 2 \,
\end{equation}
provided $\sum_i^n x_i \equiv 0 \mypmod 2$.
We say that $x=x_1x_2 \ldots x_n$ is the \emph{question} and 
$y=y_1y_2 \ldots y_n$ is the \emph{answer}.

\begin{theorem}  \label{thm:quant}
If the $n$ players are allowed to share prior entanglement, then  
they can always win game $G_n$. 
\end{theorem}

\begin{proof}
(In this proof, we assume that the reader is familiar with basic
concepts of quantum information processing~\cite{chuang}.)
Define the following $n$-qubit entangled quantum states \ket{\Phi_n^+} and
\ket{\Phi_n^-}.
\begin{align} 
\ket{\Phi_n^+} &= \oosrt \ket{0^n} + \oosrt \ket{1^n} \nonumber\\[1ex]
\ket{\Phi_n^-} &= \oosrt \ket{0^n} - \oosrt \ket{1^n} \, . \nonumber
\end{align}
Let $H$ denote the Walsh-Hadamard transform, defined as usual by
\begin{align}
H \ket{0} &\mapsto \oosrt\ket{0}+\oosrt\ket{1} \nonumber\\
H \ket{1} &\mapsto \oosrt\ket{0}-\oosrt\ket{1} \nonumber 
\end{align}
and let $S$ denote the unitary transformation defined by 
\begin{align}
S \ket{0} &\mapsto \phantom{i}\ket{0} \nonumber\\
S \ket{1} &\mapsto i\ket{1} \, . \nonumber
\end{align}


It is easy to see that if $S$ is applied to any two qubits of \ket{\Phi_n^+},
while the other qubits are left undisturbed, then the resulting state is
\ket{\Phi_n^-}, and if $S$ is applied to any two qubits of \ket{\Phi_n^-},
then the resulting state is \ket{\Phi_n^+}.  Therefore, if the qubits of
\ket{\Phi_n^+} are distributed among the $n$ players, and if exactly $m$ of them
apply $S$ to their qubit, the resulting global state will be
\ket{\Phi_n^+} if \mbox{$m \equiv 0 \mypmod4$} and
\ket{\Phi_n^-} if \mbox{$m \equiv 2 \mypmod4$}.

Moreover, the effect of applying the Walsh-Hadamard transform to each qubit in
\ket{\Phi_n^+} is to produce an equal superposition of all classical $n$-bit
strings that contain an even number of~1s, whereas the effect of applying
the Walsh-Hadamard transform to each qubit in
\ket{\Phi_n^-} is to produce an equal superposition of all classical $n$-bit
strings that contain an odd number of~1s. More formally,
\begin{align}
(H^{\cross n})  \ket{\Phi_n^+} &= 
{\textstyle \frac{1}{\sqrt{2^{n-1}}}} \!\!
 \sum_{\substack{\Delta(y)\equiv 0 \\\mypmod 2}} \ket{y} \nonumber \\
(H^{\cross n})  \ket{\Phi_n^-} &= 
{\textstyle \frac{1}{\sqrt{2^{n-1}}}} \!\!
 \sum_{\substack{\Delta(y)\equiv 1 \\\mypmod 2}} \ket{y} \, , \nonumber 
\end{align} 
where $\Delta(y) = \sum_i y_i$ denotes the Hamming weight of~$y$.

The quantum winning strategy should now be obvious.
In~the initialization phase, state \ket{\Phi_n^+} is produced and its
$n$ qubits are distributed among the $n$ players.
After they have moved apart, each player $A_i$ receives input
bit~$x_i$ and does the following.
\begin{enumerate}
\item\label{stepone} If $x_i = 1$, $A_i$ applies transformation $S$ to his qubit; 
  otherwise he does nothing.
\item\label{steptwo} He applies $H$ to his qubit.
\item He measures his qubit in order to obtain $y_i$.
\item He produces $y_i$ as his output.
\end{enumerate}

We know by the promise that an even number of players will apply $S$ to
their qubit.  If that number is divisible by~4, which means that
$\squash{\frac12} {\sum_i^n x_i}$ is even, then the global
state reverts to \ket{\Phi_n^+} after step~\ref{stepone} and therefore to
a superposition of all \ket{y} such that \mbox{$\Delta(y) \equiv 0 \mypmod2$}
after step~\ref{steptwo}.  It~follows that $\sum_i^n y_i$, the number of players
who measure and output~1, is even. On~the other hand, if
the number of players who apply $S$ to their qubit is congruent to 2
modulo~4, which means that $\squash{\frac12} {\sum_i^n x_i}$ is odd, then the
global state evolves to \ket{\Phi_n^-} after step~\ref{stepone} and
therefore to a superposition of all \ket{y} such that \mbox{$\Delta(y) \equiv 1
\mypmod2$} after step~\ref{steptwo}.  It~follows in this case that
$\sum_i^n y_i$ 
is odd.
In~either case, Equation~(\ref{goal}) is fulfilled at the end of the protocol,
as required. 
\qed
\end{proof}

\section{Optimal Proportion for Deterministic Protocols}
\label{classical}

In this section, we study the case of \emph{deterministic} classical
protocols to play game $G_n$.
We~show that no such protocol can succeed on a proportion
of the allowed inputs that is significantly better than~\pbfrac12.

\begin{theorem} \label{thm:prop}
The best possible deterministic strategy for game $G_n$ is successful
in proportion $\optprob$. 
\end{theorem}

\begin{proof}
Since no information may be communicated between players during the game, 
the best they can do is to agree on a strategy before 
the game starts.  Any such deterministic strategy will be such that
player $A_i$'s answer $y_i$ \mbox{depends} only on his input bit~$x_i$.  
Therefore, each  player has an individual strategy 
\mbox{$s_i \in \{01,10,00,11 \}$}, 
where the first bit of the pair denotes the strategy's output $y_i$ if the input
bit is \mbox{$x_i=0$} and the second bit of the strategy denotes its output if
the input is \mbox{$x_i=1$}.  In~other words, $00$ and $11$ denote the two
constant strategies $y_i=0$ and $y_i=1$, respectively,
$01$ denotes the strategy that sets \mbox{$y_i=x_i$}, and $10$ denotes the
complementary strategy \mbox{$y_i=\overline{x_i}$}.

Let $s=s_1, s_2,\ldots, s_n$ be the global deterministic strategy 
chosen by the players.
The order of the players is not important, so that
we may assume without loss of generality that 
strategy $s$ has the following form.

\[
 s= \overbrace{01, 01, \ldots , 01}^{k-\ell},
\overbrace{10, 10, \ldots , 10}^{\ell},
 \overbrace{00, 00, \ldots , 00}^{n-k-m},
 \overbrace{11, 11, \ldots , 11}^{m}
\]

Assuming strategy $s$ is being used, the Hamming weight $\Delta(y)$ of the answer
is given by

\vspace{-1ex}
\begin{align}
\Delta(y) 
     & =  
       \Delta(x_{1} \ldots, x_{k-\ell})  +
       \Delta(\overline{x_{k-\ell+1}}, \ldots, \overline{x_k}) \notag + 
       \Delta(\,\overbrace{00 \ldots 0}^{n-k-m}\,) + 
       \Delta(\,\overbrace{11 \ldots 1}^{m}\,)   \\
     &\equiv \Delta(x_{1}, \ldots, x_k) + \ell + m \pmod 2 \, . \nonumber
\end{align}
Consider the following four sets, for \mbox{$a,b \in \{0,1\}$}.
\[
S_{a,b}^k = 
\{x \ | \ \Delta(x_{1}, \ldots ,x_k) \equiv a \mypmod 2 \ \textup{and} \ %
\Delta(x_{1}, \ldots ,x_n) \equiv 2b \mypmod 4 \}
\]
If  $\ell+m$ is even then there are exactly $\smash{|S^k_{0,0}|+ |S^k_{1,1}|}$ 
questions that yield a winning answer, and otherwise if 
 $\ell+m$ is odd then there are exactly  $\smash{|S^k_{1,0}|+ |S^k_{0,1}|}$ 
questions that yield a winning answer. We also have that 
the four sets account for all possible questions and therefore
\[
 |S^k_{0,0}|+ |S^k_{1,1}| \ = \ 2^{n-1} - ( |S^k_{1,0}|+ |S^k_{0,1}|)  \, .
\]
>From here, the proof of the Theorem follows directly from
Lemma~\ref{lem:sum} below.\qed
\end{proof}

\begin{samepage}

\noindent
First we need to state a standard Lemma.
\begin{lemma}\textup{\cite[Eqn.~1.54]{Gould}}
\label{lem:binomial}
\begin{equation}
 \sum_{\substack{i \equiv a \\ \mypmod 4}}
\!\!\binom{n}{i}~=~
\begin{cases}
2^{n-2} + 2^{\frac{n}{2}-1}  &\textup{if}~n-2a \equiv 0  \mypmod 8   \\
2^{n-2} - 2^{\frac{n}{2}-1}  &\textup{if}~n-2a \equiv 4  \mypmod 8 \\
2^{n-2}                     &\textup{if}~n-2a \equiv 2,6   \mypmod 8  \\
2^{n-2}+2^{\frac{n-3}{2}}  &\textup{if}~n-2a \equiv 1,7   \mypmod 8  \\
2^{n-2}-2^{\frac{n-3}{2}} &\textup{if}~n-2a \equiv 3,5   \mypmod 8 
\end{cases} 
\end{equation}
\end{lemma}

\end{samepage}

\begin{lemma}
\label{lem:sum}
If $n$ is odd, then
\[
 |S^k_{0,0}|+ |S^k_{1,1}| =
 \begin{cases}
 2^{n-2} + 2^{\frac{n-3}{2}}  &\textup{if}~ (n-1)/2+3(n-k) \equiv 0,3 \mypmod 4\\
 2^{n-2} - 2^{\frac{n-3}{2}}  &\textup{if}~ (n-1)/2+3(n-k) \equiv 1,2 \mypmod 4
\end{cases}
\]
On the other hand, if $n$ is even, then
\[
 |S^k_{0,0}|+ |S^k_{1,1}| =
 \begin{cases}
 2^{n-2}                      &\textup{if}~n/2+3(n-k) \equiv 1,3  \mypmod 4 \\
 2^{n-2} + 2^{\frac{n}{2}-1}  &\textup{if}~n/2+3(n-k) \equiv 0 \mypmod 4\\
 2^{n-2} - 2^{\frac{n}{2}-1}  &\textup{if}~n/2+3(n-k) \equiv 2 \mypmod 4
\end{cases}
\]

\end{lemma}

\begin{proof}

\mbox{}From the definition of $S^k_{a,b}$, provided we consider that
$\binom{0}{\mbox{\footnotesize $a$}}=0$ whenever $a \neq 0 $ and $\binom{0}{0} =1$,
we get
\begin{align}
\label{eqn:1}
&|S^k_{0,0}| = 
\! \!\! \!  \sum_{\substack{i \equiv 0 \\ \mypmod 4}} \!\! \binom{k}{i} 
\! \! \! \!     \sum_{\substack{j \equiv 0 \\ \mypmod 4}} \!\!  \binom{n-k}{j} +
\! \! \! \!  \sum_{\substack{i \equiv 2 \\ \mypmod 4}} \!\! \binom{k}{i} 
\! \! \! \!    \sum_{\substack{j \equiv 2 \\  \mypmod 4}} \!\! \binom{n-k}{j}
\\[1ex]
\label{eqn:2}
&|S^k_{1,1}| =
\! \! \! \! \sum_{\substack{i \equiv 1 \\ \mypmod 4}} \!\! \binom{k}{i} 
\! \! \! \!     \sum_{\substack{j \equiv 1 \\ \mypmod 4}} \!\! \binom{n-k}{j} +
\! \! \! \! \sum_{\substack{i \equiv 3 \\ \mypmod 4}} \!\! \binom{k}{i} 
\! \! \! \!   \sum_{\substack{j \equiv 3 \\ \mypmod 4}} \!\! \binom{n-k}{j} \, .
\end{align}

Using Lemma \ref{lem:binomial}, we compute 
(\ref{eqn:1}) and (\ref{eqn:2}).  
Since $n$ and~$k$ are parameters for the equations, 
and since Lemma \ref{lem:binomial} depends on the
values of $n$ and $k$ modulo~8, we have $8$ cases to verify for $n$ and 
$8$ cases for~$k$, hence 64 cases in total.
These straightforward, albeit tedious, calculations are left to the reader.
\qed
\end{proof}

\begin{theorem} \label{thm:achievable}
Very simple deterministic protocols achieve the bound given in
Theorem~\ref{thm:prop}. In~particular, the players do not even have to look
at their input when \mbox{$n \not\equiv 2 \mypmod 4$}!
\end{theorem}

\begin{proof}
The following simple strategies, which depend on $n \mypmod 8$, are
easily seen to succeed in proportion exactly~$\optprob$.
They are therefore optimal among all possible deterministic classical strategies.
\qed\end{proof}
\begin{table}[ht]
\caption{\label{table:win}Simple optimal strategies.}
\centering
\begin{tabular}{|c|c|c|}\hline
 $n \mypmod 8$ & player 1  & players 2 to $n$    \\ \hline   
  0   &  00      &        00       \\ \hline  
  1   &  00      &        00       \\ \hline  
  2   &  01      &        00       \\ \hline  
  3   &  11      &        11       \\ \hline  
  4   &  11      &        00       \\ \hline  
  5   &  00      &        00       \\ \hline  
  6   &  10      &        00       \\ \hline    
  7   &  11      &        11       \\ \hline  
\end{tabular}
\end{table}

\section{Optimal Probability for Classical Protocols}
\label{probabilistic}

In this section, we consider all possible \emph{classical}
protocols to play game $G_n$, \mbox{including} probabilistic protocols.
We~give as much power as possible to the classical model
by allowing the playing parties unlimited sharing of random variables.
Despite this, we prove that no classical protocol can succeed
with a probability that is significantly better than~\pbfrac{1}{2}
on the worst-case input.

\begin{definition}\label{strategy}
A probabilistic strategy is a probability distribution over 
a set of deterministic strategies.
\end{definition}
The random variable shared by the players during the initialization phase
\mbox{corresponds} to deciding which deterministic strategy will be used for any
given run of the protocol.

\begin{lemma} \label{thm:yao} 
Consider any multi-party game of the sort formalized in Section~\ref{introduction}.
For any probabilistic protocol that is successful with probability~$p$,
there exists a deterministic protocol that is successful in proportion
at least~$p$.
\end{lemma}

\begin{proof}
This Lemma is a special case of a theorem proven by Andrew Yao~\cite{Y77},
but its proof is so simple that we include it here for completeness.
Consider any probabilistic strategy that is successful
with probability~$p$.
Recall that this means that the protocol wins the game with probability
at least $p$ on any instance of the problem that satisfies the promise.
By~the pigeon hole principle, the same strategy wins the game with probability
at least $p$ if the input is chosen uniformly at random among all possible
inputs that satisfy the promise.  In~other words, it is successful
\emph{in proportion} at least~$p$.  Consider now the
deterministic strategies that enter the definition of our probabilistic
strategy, according to Definition~\ref{strategy}.  Assume for a contradiction that
the best among them succeeds in proportion \mbox{$q<p$}.
Then, again by the pigeon hole principle, any probabilistic mixture of
those deterministic strategies (not only the uniform mixture) would succeed
in proportion no better than~$q$.  But this includes the probabilistic
strategy whose existence we assumed, which does succeed in proportion at
least~$p$.  This implies that \mbox{$p \le q$}, a contradiction, and therefore
at least one deterministic strategy must succeed in proportion at least~$p$.\qed
\end{proof}

\begin{samepage}

\begin{theorem} \label{thm:prob}
No classical strategy for game $G_n$ can be successful
with a probability better than $\optprob$.
\end{theorem}

\begin{proof}
Any classical strategy for game $G_n$ that would be successful with
probability \mbox{$p>\optprob$} would imply by Lemma~\ref{thm:yao}
the existence of a deterministic strategy that would succeed in proportion
at least~$p$.  This would contradict \mbox{Theorem}~\ref{thm:prop}.\qed
\end{proof}

\end{samepage}

Theorem~\ref{thm:prob} gives an upper bound on the best probability
that can be achieved by any classical strategy in winning game~$G_n$.
However, it is still unknown if there exists a classical strategy
capable of succeeding with probability~$\optprob$.
We~conjecture that this is the case.
Consider the probabilistic strategy that chooses uniformly at random among all
the deterministic strategies that are \mbox{optimal} according to
Theorem~\ref{thm:prop}. We~have been able to prove with the help of Mathematica
that this probabilistic strategy is successful with probability~$\optprob$ for all
\mbox{$3 \le n \le 14$}. We~have also proved that this probabilistic strategy is
successful with probability~$\optprob$ for any odd number $n$ of players, but only
when the players all receive \mbox{$x_i=0$} as input.
The general case is still open.

\begin{conjecture}\label{conj}
There is a classical strategy for game $G_n$ that is successful
with a probability that is \emph{exactly} $\optprob$ on all inputs.
\end{conjecture}

\section{Imperfect Apparatus}
\label{loophole}

Quantum devices are often unreliable and thus we cannot expect to 
witness the perfect result predicted by quantum mechanics
in Theorem~\ref{thm:quant}.
However, the \mbox{following} analysis shows that a reasonably large error 
probability can be tolerated if we are satisfied with making experiments
in which a quantum-mechanical strategy will succeed with a probability
that is still better than anything classically achievable.
This would be sufficient to rule out classical theories of the universe.

First consider the following model of imperfect apparatus.
Assume that the classical bit $y_i$ that is output by each  
player $A_i$ corresponds to the predictions of quantum mechanics
(if~the apparatus were perfect) with some probability~$p$.
With complementary probability \mbox{$1-p$}, the player would
output the complement of that bit.
Assume furthermore that the errors are independent between players.
In other words, we model this imperfection by saying that each player flips his
(perfect) output bit with probability \mbox{$1-p$}. 

\begin{theorem}\label{BSC}
For all $p > \squash{\frac{1}{2}}+ \squash{\frac{\sqrt{2}}{4}} \approx 85\%$
and for all sufficiently large number $n$ of players, provided each player outputs
what is predicted by quantum mechanics (according to the protocol given in the
proof of Theorem~\ref{thm:quant}) with probability at least~$p$, the quantum
success probability in  game $G_n$ remains strictly greater than anything
classically achievable.   
\end{theorem}

\begin{proof}
In the $n$-player imperfect quantum protocol, the probability $p_n$ that the game
is won is given by the probability 
of having an even number of errors:
\[
 p_n  ~ =  \sum_{\substack{i \equiv 0 \\ \mypmod 2}} \!\!
 \binom {n} {i} p^{n-i} (1-p)^{i}  \, .
\]
It is easy to prove by mathematical induction that 
\[
 p_n \ = \ \frac{1}{2} + \frac{(2p-1)^n}{2} \, .
\]
Let's concentrate for now on the case where $n$ is odd.
By Theorem \ref{thm:prob}, the success probability of any classical protocol 
is upper-bounded by 
\[ p'_n =  \frac{1}{2} + \frac{1}{2^{(n+1)/2}} \, .\] 
%
%
For any fixed $n$, define
\[ e_n = \frac{1}{2} + \frac{(\sqrt2\,)^{1+1/n}}{4} \, . \]
It follows from elementary algebra that 
\[
 p > e_n ~\Rightarrow~
 p_n  > p'_n  \, .
\]
In other words, the imperfect quantum protocol on $n$ players surpasses
anything classically achievable provided \mbox{$p > e_n$}.
For example, $e_3 \approx 89.7\%$ and $e_5 \approx 87.9\%$.
Thus we see that even the game with as few as 3 players is sufficient
to \mbox{exhibit} genuine quantum behaviour if the apparatus is at least 90\%
reliable. As~$n$ \mbox{increases}, the threshold $e_n$ decreases. In~the limit of
large $n$, we have 
\[ \lim_{n \rightarrow \infty}  e_n = \frac{1}{2}+ \frac{\sqrt{2}}{4}
\approx 85\% \, .  \]
The same limit is obtained for the case when $n$ is even.\qed
\end{proof}

Another way of modelling the imperfect apparatus is to assume that it
gives the correct answer most of the time, but sometimes it fails
to give any answer at all.  This is the type of behaviour that gives rise to
the infamous \emph{detection loophole} in experimental tests of the fact
that the world is not classical~\cite{massar}.
When the detectors fail to give an answer, the corresponding
player knows that all information is lost.
In~this case, he has nothing better to do than output a random bit.
With this strategy, either every player is lucky enough to register
an answer, in which case the game is won with certainty,
or at least one player outputs a random answer, in which case
the game is won with probability~\pbfrac12 regardless of what the
other players~do.

\begin{corollary}
For all $q > \oosrt \approx 71\%$
and for all sufficiently large number $n$ of players,
provided each player outputs what is predicted by quantum mechanics
(according to the protocol given in the proof of Theorem~\ref{thm:quant})
when he receives an answer from his apparatus with probability at least $q$,
but otherwise the player outputs a random answer,  the data collected
in playing game $G_n$ cannot be \mbox{explained} by any classical local realistic
theory.
\end{corollary}

\begin{proof}
If a player obtains the correct answer with probability $q$
and otherwise outputs a random answer, the probability that the resulting
output be correct is \mbox{$p=q+\smash{\frac12}(1-q)=(1+q)/2$}.
Therefore, this scenario reduces to the previous one with this simple
change of variables.
We know from Theorem~\ref{BSC} that the imperfect quantum protocol
is more reliable than any possible classical protocol, provided $n$ is
large enough, when \mbox{$p > \squash{\frac{1}{2}}+ \squash{\frac{\sqrt{2}}{4}}$}.
This translates directly to $q>\oosrt$.\qed
%
\end{proof}

\section{Conclusions and Open Problems}

We have demonstrated that quantum pseudo-telepathy can arise for simple
multi-party problems that cannot be handled by classical protocols much better
than by the toss of a coin.  This could serve to design new tests for the
nonlocality of the physical world in which we live.

In closing, we propose two open problems.
First, can Conjecture~\ref{conj} be proven or are the best possible classical
probabilistic protocols for our game even worse than hinted at by
Theorem~\ref{thm:prob}? Second, it would be nice to find a \emph{two}-party
pseudo-telepathy problem that admits a perfect quantum solution, yet any
\mbox{classical} \mbox{protocol} would have a small probability
of success even for inputs of small or moderate size.


\begin{thebibliography}{[99]}


\bibitem{aaronson}
Aaronson,~S., Ambainis,~A.:
Quantum search of spatial regions.
Available as \texttt{arXiv:quant-ph/0303041} (2003).

\bibitem{teleport}
Bennett,~C.\,H., Brassard,~G., Cr\'epeau,~C., Jozsa,~R., Peres,~A.,
Wootters,~W.\,K.:
Teleporting an unknown quantum state via dual classical and
{E}instein--{P}odolsky--{R}osen channels.
\textit{Physical Review Letters} \textbf{70} (1993) 1895\,--\,1899.

\bibitem{survey}
Brassard,~G.:
Quantum communication complexity.
{\em Foundations of Physics}
(to~appear, 2003).

\bibitem{BCT99}
Brassard,~G., Cleve,~R., Tapp,~A.:
Cost of exactly simulating quantum entanglement 
with classical  communication. 
{\em Physical Review Letter} \textbf{83} (1878) 1874\,--\,1878. 

\bibitem{BCW}
Buhrman,~H., Cleve,~R., Wigderson,~A.:
Quantum vs.\ classical communication and computation.
\textit{Proceedings of 30th Annual ACM Symposium on Theory of Computing}
(1998) 63\,--\,68. 


\bibitem{CB}
Cleve, R., Buhrman, H.:
Substituting quantum entanglement for communication.
\textit{Physical Review~A}
\textbf{56} (1997) 1201\,--\,1204.



\bibitem{Gould}
Gould, H.\,W.:
\emph{Combinatorial Identities}.
Morgantown (1972).

\bibitem{KS}
Kalyanasundaram, B., Schnitger, G.:
The~probabilistic communication complexity of set intersection.
\textit{Proceedings of 2nd Annual IEEE Conference on Structure in
Complexity Theory}
(1987) 41\,--\,47. 


\bibitem{massar}
Massar, S.:
Non locality, closing the detection loophole, and communication complexity.
\textit{Physical Review~A} \textbf{65} (2002) 032121-1\,--\,032121-5.

\bibitem{chuang}
Nielsen, M.\,A., Chuang, I.\,L.:
\emph{Quantum Computation and Quantum Information}.
Cambridge University Press (2000).

\bibitem{raz}
Raz, R.:
Exponential separation of quantum and classical communication complexity.
\textit{Proceedings of 31st Annual ACM Symposium on Theory
of Computing}
(1999) 358\,--\,367. 

\bibitem{Y77}
Yao, A.\,C.--C.:
Probabilistic computations: Toward a unified measure of complexity.
\textit{Proceedings of 18th IEEE Symposium on Foundations of Computer Science}
(1977) 222\,--\,227.

\bibitem{yao83}
Yao, A.\,C.--C.:
Quantum circuit complexity.
\textit{Proceedings of 34th Annual IEEE \mbox{Symposium} on Foundations of
Computer Science}
(1993) 352\,--\,361. 



\end{thebibliography}

\bibliographystyle{splncs}

\end{document}